\documentclass{ieeeaccess}
\usepackage{cite}
\usepackage{amsmath,amssymb,amsfonts}
\usepackage{algorithmic}
\usepackage{graphicx}
\usepackage{textcomp}
\def\BibTeX{{\rm B\kern-.05em{\sc i\kern-.025em b}\kern-.08em
    T\kern-.1667em\lower.7ex\hbox{E}\kern-.125emX}}
\begin{document}
\history{Date of publication xxxx 00, 0000, date of current version xxxx 00, 0000.}
\doi{}

\title{Design and Analysis of Passband Transmitted Reference Pulse Cluster UWB Systems in the Presence of Phase Noise}
\author{\uppercase{Zhonghua Liang}\authorrefmark{1}, \IEEEmembership{Member, IEEE},
\uppercase{Guowei Zhang\authorrefmark{2}, Xiaodai Dong\authorrefmark{2}}, \IEEEmembership{Senior Member, IEEE}, \uppercase{and Yiming Huo\authorrefmark{2}},
\IEEEmembership{Member, IEEE}}
\address[1]{School of Information Engineering, Chang'an University, Xi'an 710064, China}
\address[2]{Department of Electrical and Computer Engineering, University of Victoria, Victoria, BC V8P 5C2, Canada}

\tfootnote{The work of Zhonghua Liang was supported in part by the National Natural Science Foundation of China (NSFC) under Grant 61271262 and 61572083, and in part by the Natural Science Basic Research Plan in Shaanxi Province of China under Grant 2017JM6099, and by the Fundamental Research Funds for the Central Universities of China under Grant 310824171004.}

\markboth
{Liang \headeretal: Design and Analysis of Passband TRPC-UWB Systems in the Presence of Phase Noise}
{Liang \headeretal: Design and Analysis of Passband TRPC-UWB Systems in the Presence of Phase Noise}

\corresp{Corresponding author: Xiaodai Dong (e-mail: xdong@ece.uvic.ca).}

\begin{abstract}
Transmitted reference pulse cluster (TRPC) signaling was recently proposed and 
developed for noncoherent ultra-wideband (UWB) communications. In this paper, 
a practical passband TRPC-UWB system is designed and analyzed to deal with the 
carrier frequency offset, phase offset and phase noise inherent in voltage-controlled 
oscillators (VCO) of the transmitter and the receiver. Based on a general model of noisy 
VCO and employing some reasonable assumptions, an equivalent linear time-invariant (LTI) 
analytical model is obtained to facilitate the bit error rate (BER) analysis. Our analysis 
shows that the constant carrier frequency offset and the phase offset can be removed by 
employing the passband transmitter and the noncoherent receiver. Furthermore, a semi-analytical 
BER expression is derived to show the impact of phase noise on the system error performance. 
Simulation results validate the semi-analytical expressions and both of them indicate that TRPC 
is more robust to the effect of phase noise than conventional transmitted reference (TR) and 
coherent UWB Rake receivers.\end{abstract}


\titlepgskip=-15pt

\maketitle

\section{Introduction}\label{sec1}
\PARstart{S}{ince} transmitted reference (TR) signaling was first introduced to ultra-wideband (UWB) communications in 2002\cite{Hoctor_Delay,Choi_Performance}, this technique has attracted considerable interest due to its simplicity and robust performance as a noncoherent impulse radio technology\cite{D'Amico_GLRT,Chao_UWB,Quek_Analysis,Romme_wacr_tr,Quek_Unified_Analysis,Niranjayan_TR,D'Amico_TR_UWB}. The autocorrelation receiver (AcR) of a TR system does not need to do explicit channel estimation, which is particularly challenging for UWB channels, and only low sampling rate is used at the analog output of AcR, avoiding the need for high rate analog to digital conversion  of a UWB signal. Presently, TR in conjunction with AcR and binary pulse position modulation (BPPM) signaling combined with energy detector (ED) have become two popular noncoherent UWB systems\cite{Witrisal_UWB}.

One major challenge to the development of TR-UWB systems is the need to implement long ultra-wideband delay lines in AcR, which is not practically feasible\cite{Casu_Implementation,VanStralen_Delay,D'Amico_TR_UWB}. Accordingly, a variety of approaches were proposed to address this implementation issue from time, frequency, and code domains, respectively\cite{Goeckel_FSR,Dong_trpc,D'Amico_CMR}. Among them, the time-domain solution\cite{Dong_trpc} is referred to as transmitted reference pulse cluster (TRPC) signaling, where compactly and uniformly placed multiple pulses forming a
pulse cluster are used and therefore only very short delay lines are required. Moreover, TRPC exhibits significant advantages over the original TR in terms of bit error rate (BER) performance and data rate due to the compact pulse structure. Hence it is a promising candidate for a wide range of impulse radio applications.

Recently, the TRPC-UWB system was further developed and improved from various practical aspects\cite{Jin_Int_Det_TWC,Liang_MA_TRPC,Liang_AD_TRPC,Jin_iTRPC,Liang_FEC_TRPC,Dai_trpc_ppm_bpsk,Jin_TRPC_Digi_Rec,Huo_TX_UWB_Chip,Dai_trpc_coop_relay,Liang_IDT_iTRPC,Liang_LDPC_TRPC,Jin_TRPC_AVG_Rec,Huo_TX_UWB}. All of these research results are focused on the so-called baseband or carrierless TRPC-UWB, where the information-bearing burst pulses are directly radiated at the transmit antenna. Actually, most impulse radio systems studied in the literature are carrierless because baseband or carrierless transmission implies that an impulse radio may be manufactured inexpensively\cite{Win_IR_UWB}. 

However, the carrierless form that makes impulse radio attractive also results in some design challenges. First of all, judicious pulse shaping or designing algorithms need to be developed to guarantee that the transmit spectra meet Federal Communications Commission (FCC) spectral mask constraints inside the frequency band of 3.1-10.6 GHz assigned to UWB devices\cite{FCC_Part15}. Although various useful pulse shapes were proposed to satisfy the FCC requirements\cite{Hu_Pulse,Bai_UWB_pulse,Kim_Orth_pulse}, most of them either have high implementation complexity, or lack the flexibility in power spectral density (PSD) fitting processes, or are feasible only in theory and yet to be demonstrated in practical applications for digital circuitry implementation.

On the other hand, based on the FCC mask regulations, the IEEE 802.15.4a standard allocates sixteen operating frequency bands for its alternative UWB physical layer (PHY), each of which (except band 0) occupies a bandwidth of at least 499.2 MHz centered on a center frequency that ranges from 3.5-10 GHz\cite{IEEE802154a_standard}. Hence, it indicates that carrier implementation can also be considered to provide the flexibility of hopping among the defined operating frequency bands, because this flexibility can enhance the coexistence of UWB PHYs with other wireless devices operating in the same spectrum, and it can be helpful to increase the coordinated piconet capabilities for simultaneously operating UWB piconets as well. 

Motivated by the discussions mentioned above, a passband TRPC-UWB system is proposed in this paper aiming for practical implementation of low cost, low power consumption and low complexity systems. In the passband TRPC-UWB system, the information-bearing baseband pulses that occupy a bandwidth of about 500 MHz are up-converted to gigahertz bands and transmitted through the antenna. Compared to the carrier-free form, the passband TRPC-UWB can flexibly switch its operating frequency band by simply changing the carrier center frequency and with the same baseband pulse shape.

In practice, voltage-controlled oscillators (VCO's) are essential building blocks for both wideband radar and carrier communication systems. All practical oscillators have phase noise components, and it is well known that phase noise degrades the system performance severely if not dealt with properly\cite{Razavi_phase_noise,Demir_phase_noise,Demir_phase_noise_jitter,Wilcoxson_Master,Nouri_PN_Radar}. Moreover, the time-varying perturbation effect of phase noise has been well investigated in extensive literature on circuit theory, circuit systems and applications since 1960's\cite{Lax_classical_noise,Kaertner_corr_spectrum,Kaertner_analysis_noise,Hajimiri_PN,Demir_phase_noise,Demir_phase_noise_jitter,Ham_virtual_damping,Li_phase_noise}, where the Brownian motion process has been accepted as a general model to characterize and analyze the phase noise. Wiener or Brownian motion process is the general model to describe the phase noise of a free-running oscillator. In conventional transceivers, relatively expensive phase locked loop (PLL) frequency synthesizers are much more widely used than free-running oscillators. The PLL noise is usually modeled as a multidimensional Ornstein-Uhlenbeck process \cite{Mehrotra_pll_noise}. However, if the passband TRPC transceiver allows the use of a simple, low cost free running VCO, the system cost will be further reduced and this is significant for a general class of applications with low-cost and low-power consumption. Hence, only the Brownian motion process deriving from free-running oscillators is considered in this paper.

To the best of our knowledge, a comprehensive and detailed system performance analysis on phase noise is not yet available for \textit{impulse radio} in the literature so far. This is partly because most impulse radios in the literature consider ``carrier-free''. Even for other carrier-based UWB systems, such as single carrier direct sequence (SC-DS-) UWB and multiband orthogonal frequency division multiplexing (MB-OFDM-) UWB, only a few works were reported \cite{Razavi_UWB_COMS,Park_DS_CMOS} and they were focused on some specifications for the implementation of CMOS transceivers based on the measured phase noise profiles of VCO's.

On the other hand, for conventional OFDM systems in which phase noise causes common phase error and intercarrier interference, performance analysis and phase noise mitigation algorithms have been extensively investigated\cite{Petrovic_PN_OFDM,Zou_PN_OFDM,Lee_PN_OFDM}. Nevertheless, these results cannot be easily employed in impulse radios because of the significant differences between impulse radio and conventional OFDM systems.

Based on the widely-accepted model of noisy oscillator reported in\cite{Demir_phase_noise,Demir_phase_noise_jitter,Wilcoxson_Master,Lax_classical_noise,Kaertner_corr_spectrum,Kaertner_analysis_noise,Hajimiri_PN,Ham_virtual_damping,Li_phase_noise}, our previous work\cite{Liang_phase_noise_vtc} analyzed the phase noise effect on the BER performance for the TRPC-UWB system. However, it did not include carrier frequency offset, which is a major practical issue if free-running VCO's are used. Hence, this paper presents a theoretical framework about the design and analysis of passband TRPC-UWB systems in the presence of phase noise, carrier frequency offset, and phase offset. In our analysis, the phase noise is modeled as a Brownian motion process, and the constant carrier frequency offset and phase offset between the transmitter and the receiver are considered as uniformly distributed random variables (r.v.'s). With the passband TRPC transmitter and the noncoherent receiver structure, the derivation of the bit error rate (BER) performance is given in detailed and it shows that the constant carrier frequency offset and the phase offset can be cancelled. Moreover, this paper presents a thorough comparison with the conventional passband TR and the passband coherent Rake receiver on the impact of phase noise. TRPC is shown to have superior robust performance to phase noise. 

The remainder of this paper is organized as follows. Section II briefly describes the system models and assumptions for passband TRPC-UWB, including the general model of noisy oscillator and some assumptions for performance analysis. Section III presents an equivalent linear time invariant (LTI) analytical model for passband TRPC-UWB to facilitate the analysis and derives the semi-analytical BER expression. Semi-analytical and simulation results are presented and discussed in Section IV. Finally, conclusions are provided in Section V.
 
\Figure[t!](topskip=0pt, botskip=0pt, midskip=0pt)[scale=0.69]{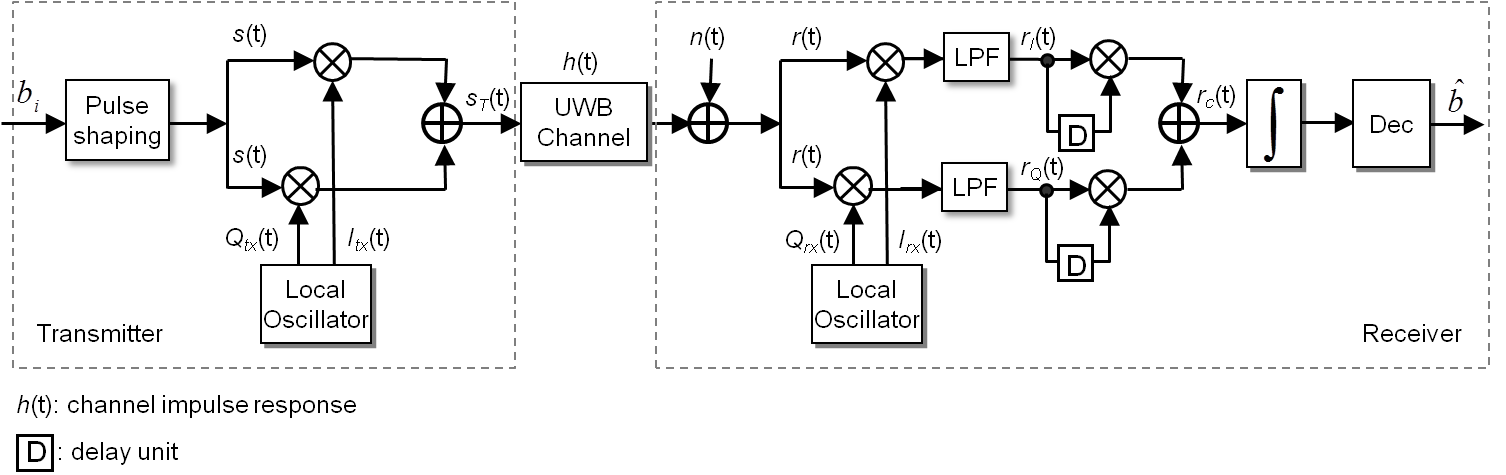}
{Block diagram of the passband TRPC-UWB system.\label{block_diagram}}

Throughout this paper $(\cdot)^T$ and $\otimes$ denote transpose and convolution, respectively. $E\{x\}$ and $Var\{x\}$ denote the mathematical expectation and the variance of r.v. $x$, respectively. $Cov\{x,y\}$ represents the covariance of two r.v.'s $x$ and $y$. $\textrm{sgn}\{x\}$ represents the sign of a real variable $x$.

\section{System Models and System Assumptions}\label{sec2}
We consider a passband TRPC-UWB transceiver that consists of baseband signaling, upconversion, downconversion, and baseband signal processing \& detection blocks. Accordingly, a block diagram of the passband TRPC-UWB transceiver is presented in Fig.~\ref{block_diagram}.
  
\subsection{Baseband Signaling}
For simplicity, we only consider the single-user scenario. For each bipolar symbol $b_i\in\{{\pm1\}}_{i=0}^{+\infty}$, a pulse pair composed of a reference pulse and a data pulse with very short delay $T_{d}$ is repeated uniformly every $2T_{d}$ seconds. After pulse shaping, the baseband TRPC signal for the $i$-th symbol can be written as\cite{Dong_trpc}
\begin{align}\label{baseband_trpc}
s_i(t)=\sqrt{\frac{E_b}{2N_f}} s_{b_{i}}(t-iT_s),
\end{align}
where $E_b$ is the average energy per symbol, $N_f$ represents the number of repeated pulse pairs per symbol, $T_s$ denotes the symbol duration, $s_{b_i}(t)\triangleq\sum_{m=0}^{N_f-1}[g(t-2mT_d)+b_{i}\cdot{g(t-(2m+1)T_d)}]$, where $g(t)$ is the normalized baseband pulse with duration $T_p$ and with bandwidth $B_0=500$ MHz.
Moreover, we let $T_{d}=T_{p}$, which means a cluster is composed of $N_f$ identical baseband pulse pairs consecutively. It is noted that besides the much smaller $T_{d}$ which indicates a feasible delay line, another significant improvement of TRPC over the original TR lies in the facts that 1) the short integration interval due to the compact cluster structure leads to much less noise collected by the autocorrelation receiver than the original TR; 2) the asymptotic ($N\to{\infty}$) energy efficiency for data detection in TRPC can be twice of that in the original TR, because only the energy in the first reference pulse is not used in the data detection\cite{Dong_trpc}. Although such a compact cluster structure also inevitably results in considerable inter-pulse interference (IPI), it has been demonstrated in \cite{Dong_trpc} that the two benefits mentioned above still significantly exceed the penalty caused by IPI. Moreover, the impact of IPI can be successfully mitigated via decision threshold optimization methods (more detailed discussions can be found in\cite{Liang_AD_TRPC}). For the sake of conciseness, the subscript $i$ in $s_i(t)$, $b_{i}$ and $s_{b_i}(t)$ will be omitted hereinafter. 

\subsection{Channel Model}
In this paper, only the phase noise caused by oscillators is investigated. Therefore, instead of using the complex-valued IEEE 802.15.4a channel model \cite{Molisch_Channel} in \cite{Liang_phase_noise_vtc} where the random phase perturbation caused by multipath propagation has been considered, we just consider the IEEE 802.15.3a real-valued multipath channel model \cite{Foerster_Channel}, in which the general form of channel impulse response can be written as 
\begin{equation}
h(t)=\sum_{k=0}^{K-1} \alpha_k \delta(t-\tau_k),
\end{equation}
where the real-valued parameters $\alpha_k$ and $\tau_k$ denote the fading coefficient and delay of the $k$-th multipath component, respectively. To facilitate the performance analysis, we consider the following two important assumptions on the UWB channel model:

\emph{Assumption I (Resolvable dense multipath channel)}: Actually, most analysis on UWB channels in the literature assumes the minimum multipath separation is the UWB pulse width, i.e., $\tau_k=kT_p$, to make the statistical analysis tractable\cite{Quek_Analysis,Quek_Unified_Analysis,Cassioli_UWB_channel,Win_UWB_channel}. In this work, we assume $\tau_k=kT_c$, where $T_c\triangleq1/f_{c}$ and $f_{c}$ is the nominal center frequency of the carrier. Since $f_{c}$ ranges from 3.5-10 GHz for low and high frequency bands \cite{IEEE802154a_standard}, $T_c$ varies between $0.1$ and $0.29$ ns. However, for a baseband UWB pulse with a bandwidth of about 500 MHz, $T_p\approx2$ ns. Hence, compared to $\tau_k=kT_p$, $\tau_k=kT_c$ corresponds to a much higher temporal resolution for dense multipath channels.    

\emph{Assumption II (Intersymbol interference (ISI) free scenario)}: Consider that the symbol duration $T_s$ is larger than the maximum channel delay $\tau_{max}$
plus the cluster length $2N_fT_{d}$, and therefore there is no ISI. Such an assumption is reasonable for low or medium date rate UWB
systems. For example, with a date rate of 1 Mbps ($T_{s}=10^{3}$ ns), $N_f=4$ and $T_{d}=2.02$ ns, so that $T_{s}\geq2N_fT_{d}+\tau_{max}$ holds for most IEEE 802.15.3a indoor channels \cite{Foerster_Channel}.

\subsection{Upconversion and Downconversion}
As shown in Fig.~\ref{block_diagram}, at the transmitter, the baseband TRPC signal $s(t)$ is upconverted to in-phase (I) and quadrature (Q) components by the cosine and sine carrier waveforms of the VCO output, $I_{tx}(t)$ and $Q_{tx}(t)$, respectively. Then the sum of I and Q components forms the combined passband TRPC signal transmitted at the antenna. At the receiver, the received signal $r(t)$ is first downconverted by the local cosine  and sine carrier waveforms $I_{rx}(t)$ and $Q_{rx}(t)$, respectively, and then after passing through two lowpass filters (LPFs), it results in two baseband components $r_{I}(t)$ and $r_{Q}(t)$. In the ideal case, $I_{tx}(t)$ and $I_{rx}(t)$ are the same and their output carrier frequencies are exactly equal to $f_c$, and so do $Q_{tx}(t)$ and $Q_{rx}(t)$. However, VCO's in practice always have frequency offset from the nominal center frequency and time varying phase noise. Moreover, there is also constant phase offset between the free-running VCO's of the transmitter and the receiver.   

Therefore, based on a practical model of noisy VCO \cite{Wilcoxson_Master,Hajimiri_PN} (equivalent or more original models can also be found in\cite{Demir_phase_noise,Demir_phase_noise_jitter,Lax_classical_noise,Kaertner_corr_spectrum,Kaertner_analysis_noise,Ham_virtual_damping,Li_phase_noise}), $I_{tx}(t)$, $I_{rx}(t)$, $Q_{tx}(t)$ and $Q_{rx}(t)$ can be modeled as follows:
\begin{equation}\label{Carrier_phase_noisy}
\begin{split}
&I_{tx}(t)=\cos\left[2\pi{f_c}t+\theta_{tx}(t)\right],\\
&Q_{tx}(t)=-\sin\left[2\pi{f_c}t+\theta_{tx}(t)\right],\\
&I_{rx}(t)=\cos\left[2\pi({f_c}+\Delta{f})t+\theta_{rx}(t)+\phi\right],\\
&Q_{rx}(t)=-\sin\left[2\pi({f_c}+\Delta{f})t+\theta_{rx}(t)+\phi\right],\\
\end{split}
\end{equation}
where $\Delta{f}$ and $\phi$ denote the constant carrier frequency offset and the initial phase difference between the two VCO's of the transmitter and the receiver, which can be considered as two r.v.'s uniformly distributed over $[-\xi,+\xi]$ MHz and $[0,2\pi)$, respectively. Based on our hardware measurements, the maximum magnitude of the relative constant carrier frequency offset of free-running VCO's is $1250$ ppm for the case that $3.5\leq{f_c}\leq4$ GHz. Accordingly, the maximum value of the absolute constant carrier frequency offset, $\xi$, can be obtained as $5$ MHz in this paper. For local oscillators using phase locked loop with VCO, the carrier frequency offset is much smaller than that of the free-running VCO's. Phase noise terms $\theta_{tx}(t)$ and $\theta_{rx}(t)$ are two independent Brownian motion processes, both of which are derived from the same random process $\theta(t)$ expressed by \cite{Wilcoxson_Master}
\begin{equation}\label{phase_noise}
\theta(t)=2\pi\int_0^t{\mu(\tau)d\tau}~(\textrm{for}~t>0),
\end{equation}
where $\mu(t)$ is a zero-mean white Gaussian noise process with a two-sided PSD of $N_{1}$ and therefore $\theta(t)$ can be considered as zero-mean Gaussian process with variance \cite{Wilcoxson_Master}
\begin{equation}\label{var_phase_noise}
Var[\theta(t)]=(2\pi)^2N_{1}t=2\pi\beta{t},
\end{equation}
where $\beta\triangleq2\pi{N_{1}}$ is used to characterize the severeness of the phase noise and it is also referred to as half-power or $3$-dB bandwidth of the noisy carrier, because the PSD of a noisy cosine or sine carrier given by (\ref{Carrier_phase_noisy}) has been shown to be a Lorentzian spectrum  with $3$-dB bandwidth of $2\pi{N_{1}}$\cite{Wilcoxson_Master}.

Using (\ref{baseband_trpc}) and (\ref{Carrier_phase_noisy}), we can write the transmitted passband TRPC signal as
\begin{equation}\label{passband_trpc}
s_T(t)=s(t)\cos\left[2\pi{f_c}t+\theta_{tx}(t)\right]-s(t)\sin\left[2\pi{f_c}t+\theta_{tx}(t)\right].
\end{equation}
From Fig.~\ref{block_diagram} and observing (\ref{passband_trpc}), it is noted that for the I-Q upconversion employed in this paper the I and Q components have the same input $s(t)$. Therefore, the upconverter can also be implemented by a single I-branch or Q-branch because (\ref{passband_trpc}) can be rewritten as $s_T(t)=\sqrt{2}s(t)\cos\left[2\pi{f_c}t+\theta_{tx}(t)+\pi/4\right]$. However, for the sake of analysis, (6) is still given in an I-Q form in this paper.

After the transmitted signal $s_{T}(t)$ passes through the UWB channel and the bandpass filter (BPF) at the antenna, the received signal $r(t)$ is given by 
\begin{align}\label{received_sig}
r(t)&=s_{T}(t)\otimes{h(t)}+n(t)\nonumber\\
&{=}\sum_{k=0}^{K-1} \alpha_k s(t-\tau_k)\Big\{\cos\left[2\pi{f_c}(t-\tau_k)+\theta_{tx}(t-\tau_k)\right]\nonumber\\
&{}-\sin\left[2\pi{f_c}(t-\tau_k)+\theta_{tx}(t-\tau_k)\right]\Big\}+n(t),
\end{align}
where $n(t)$ is the BPF-filtered complex additive white Gaussian noise (AWGN) with a one-sided PSD of $N_{0}$.

$r(t)$ can be downconverted with $I_{rx}(t)$ and $Q_{rx}(t)$ respectively by using the noncoherent detection technique\cite{Madhow_funmental_DCom}, and it results in two corresponding parts as follows:
\begin{align}\label{downconvert_ri}
\tilde{r}_I(t)&=r(t)I_{rx}(t)\nonumber\\
&=r(t)\cos\left[2\pi({f_c}+\Delta{f})t+\theta_{rx}(t)+\phi\right]
\end{align}
and
\begin{align}\label{downconvert_rq}
\tilde{r}_Q(t)&=r(t)Q_{rx}(t)\nonumber\\
&=-r(t)\sin\left[2\pi({f_c}+\Delta{f})t+\theta_{rx}(t)+\phi\right],
\end{align}
respectively. After $\tilde{r}_I(t)$ and $\tilde{r}_Q(t)$ are filtered by two LPFs with a bandwidth of $500$ MHz, two baseband components $r_{I}(t)$ and $r_{Q}(t)$ can be obtained respectively.

\Figure[t!](topskip=0pt, botskip=0pt, midskip=0pt)[scale=0.68]{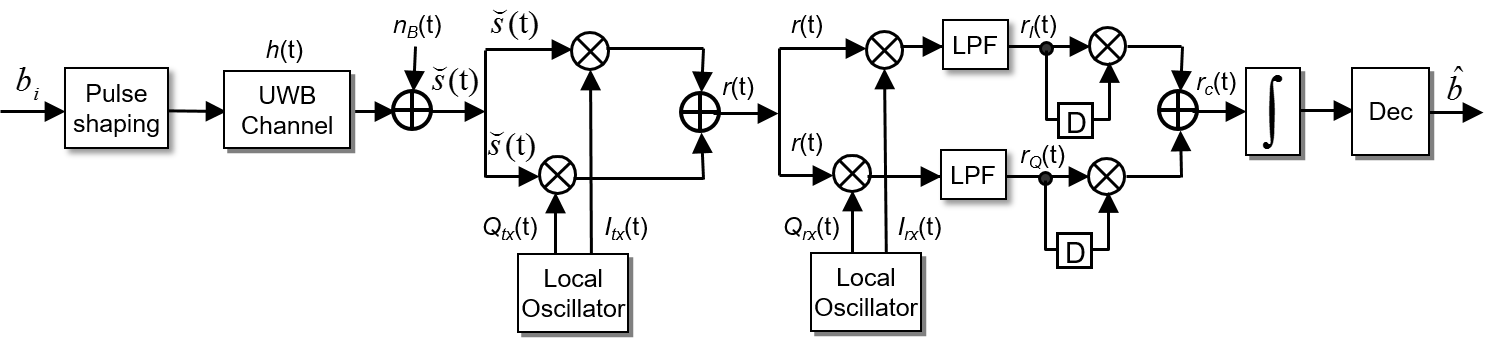}
{Equivalent LTI analytical model of the passband TRPC system.\label{equi_lti_model}} 

\subsection{Baseband Autocorrelation Processing \& Detection}
As shown in Fig.~\ref{block_diagram}, $r_{I}(t)$ and $r_{Q}(t)$ are firstly multiplied by their $T_d$ delayed version, $r_{I}(t-T_d)$ and $r_{Q}(t-T_d)$, respectively. Then the two results are combined and integrated. Therefore, the decision variable (DV) for the $i$-th symbol can be obtained as
\begin{align}\label{DV_baseband}
D=\int_{iT_{s}+T_{1}}^{iT_{s}+T_{2}}\big[\underbrace{r_{I}(t){r}_{I}(t-T_{d})+r_{Q}(t){r}_{Q}(t-T_{d}}_{\triangleq{r_{c}(t)}})\big]dt,
\end{align}
where the integration parameters $T_1$ and $T_2$ can be determined following one of the algorithms presented in \cite{Jin_Int_Det_TWC}. Finally, a symbol decision can be made as follows:
\begin{equation}\label{Decision}
\hat{b}=\textrm{sgn}\{D\}.
\end{equation}

\section{Performance Analysis in the presnece of Phase Noise}\label{sec3}
Observing (\ref{received_sig})--(\ref{downconvert_rq}), we can easily find that multiple uncertainties, $\Delta{f}$, $\phi$, $\theta_{rx}(t)$, and $\theta_{tx}(t-\tau_{k})~\textrm{for}~k=0,\cdots,K-1$, leak in the low-frequency components of $\tilde{r}_I(t)$ and $\tilde{r}_Q(t)$, and therefore, they also remain in $r_{I}(t)$ and $r_{Q}(t)$. This result is based on the sufficient but not necessary condition that $\theta(t)$ is slowly time-varying so that the maximum frequency offset from $f_c$ caused by $\theta(t)$ is much less than ${B_0}=\textrm{500~MHz}$. This condition can be verified by numerous empirical PSD profiles of phase noise, where most measured PSDs at a frequency offset of $10$ MHz are less than $-130$ dBc/Hz, which denotes the level below the detectable noise floor, for a carrier with $f_c\geq1$ GHz\cite{Berny_cmos_conf,Wu_cmos,Berny_vco_jour,Fard_PN_VCO,Huo_VCO,Huo_VCO_TRPC,Lu_ADPLL}. This makes the performance analysis intractable due to the multiple uncertainties involved. Hence, an equivalent LTI analytical model is derived in the following to facilitate the analysis. 

\subsection{Equivalent LTI Analytical Model}
Firstly, (\ref{received_sig}) can be rewritten as
\begin{align}\label{equiv_rev_trpc}
r(t)&{=}\sum_{k=0}^{K-1}\alpha_k{s(t-\tau_k)}\Big\{\cos\big[2\pi{f_c}t+\theta_{tx}(t-\tau_k)\big]\nonumber\\
&{}-\sin\big[2\pi{f_c}t+\theta_{tx}(t-\tau_k)\big]\Big\}+n(t)\nonumber\\
&{\approx}\sum_{k=0}^{K-1}\alpha_k{s(t-\tau_k)}\Big\{\cos\big[2\pi{f_c}t+\theta_{tx}(t)\big]\nonumber\\
&{}-\sin\big[2\pi{f_c}t+\theta_{tx}(t)\big]\Big\}+n(t)\nonumber\\
&{=}\breve{s}(t)\cos\big[2\pi{f_c}t+\theta_{tx}(t)\big]-\breve{s}(t)\sin\big[2\pi{f_c}t+\theta_{tx}(t)\big],
\end{align}
where the first equality holds for Assumption~I, the approximation is derived in Appendix~\ref{LTI_model}, and $\breve{s}(t)\triangleq{s(t)\otimes{h(t)}+n_B(t)}$, where $n_B(t)$ denotes the complex baseband AWGN with a one-sided PSD of $N_{0}$. According to (\ref{equiv_rev_trpc}), an equivalent LTI analytical model is shown in Fig.~\ref{equi_lti_model}.

\subsection{Derivation of BER Performance}
Based on the equivalent LTI analytical model presented in Fig.~\ref{equi_lti_model} and using (\ref{equiv_rev_trpc}), we can also rewrite (\ref{downconvert_ri}) and (\ref{downconvert_rq}) as
\begin{align}\label{equi_downconvert_ri}
\tilde{r}_I(t)&=\breve{s}(t)\Big\{\cos\big[2\pi{f_c}t+\theta_{tx}(t)\big]-\sin\big[2\pi{f_c}t+\theta_{tx}(t)\big]\Big\}\nonumber\\
&{\cdot}\cos\big[2\pi({f_c}+\Delta{f})t+\theta_{rx}(t)+\phi\big]
\end{align}
and
\begin{align}\label{equi_downconvert_rq}
\tilde{r}_Q(t)&=-\breve{s}(t)\Big\{\cos\big[2\pi{f_c}t+\theta_{tx}(t)\big]\nonumber\\
&{-}\sin\big[2\pi{f_c}t+\theta_{tx}(t)\big]\Big\}\nonumber\\
&{\cdot}\sin\big[2\pi({f_c}+\Delta{f})t+\theta_{rx}(t)+\phi\big]
\end{align}
respectively.

Employing trigonometric formulas to (\ref{equi_downconvert_ri}) and (\ref{equi_downconvert_rq}), we can easily find that only four uncertainties, namely $\theta_{tx}(t)$, $\theta_{rx}(t)$, $\Delta{f}$, and $\phi$ remain in the low-frequency terms, and the passband terms will be removed by the LPF filters. Accordingly, the two LPF-filtered baseband components are
\begin{align}\label{equi_baseband_ri}
r_I(t)&=\frac{1}{2}\breve{s}(t)\Big\{\cos\big[\Theta(t)-2\pi\Delta{f}t-\phi\big]\nonumber\\
&{-}\sin\big[\Theta(t)-2\pi\Delta{f}t-\phi\big]\Big\}
\end{align}
and
\begin{align}\label{equi_baseband_rq}
r_Q(t)&=\frac{1}{2}\breve{s}(t)\Big\{\sin\big[\Theta(t)-2\pi\Delta{f}t-\phi\big]\nonumber\\
&{+}\cos\big[\Theta(t)-2\pi\Delta{f}t-\phi\big]\Big\}
\end{align}
respectively, where $\Theta(t)\triangleq\theta_{tx}(t)-\theta_{rx}(t)$. Using (\ref{equi_baseband_ri}) and (\ref{equi_baseband_rq}) and employing trigonometric formulas, we obtain the combined baseband signal as
\begin{align}\label{combined_equi_baseband}
r_{c}(t)&=r_I(t)r_I(t-T_{d})+r_Q(t)r_Q(t-T_{d})\nonumber\\
&{=}\frac{1}{2}\breve{s}(t)\breve{s}(t-T_{d})\cos\big[\Theta(t)-\Theta(t-T_{d})-2\pi\Delta{f}{T_d}\big]\nonumber\\
&{\approx}\frac{1}{2}\breve{s}(t)\breve{s}(t-T_{d})\cos\big[\Phi(t)\big],
\end{align}
where $\Phi(t)\triangleq\Theta(t)-\Theta(t-T_{d})$, and the approximation holds because $2\pi\Delta{f}{T_d}$ can be negligible for the case that $\Delta{f}\in[-5,+5]$ MHz and $T_d=2.02$ ns. Note that the receiver step in (\ref{combined_equi_baseband}), which derives from the I-Q downconversion, can successfully cancel the constant carrier frequency offset $\Delta{f}$ and the phase offset $\phi$ between the free-running transmitter and receiver VCO's. Hence, the I-Q downconversion must necessarily be used in the noncoherent detection without the exact knowledge of carrier phase and we will also verify this point in the next section. 

Substituting (\ref{combined_equi_baseband}) into (\ref{DV_baseband}), we have
\begin{align}\label{equi_DV_baseband}
D=\frac{1}{2}\int_{iT_{s}+T_{1}}^{iT_{s}+T_{2}}\breve{s}(t)\breve{s}(t-T_{d})\cos\big[\Phi(t)\big]dt.
\end{align}
Since $\theta_{tx}(t)$ and $\theta_{rx}(t)$ are slowly time-varying random processes, $\Theta(t)$ is also a slowly time-varying random process. Hence, $\Phi(t)=\Theta(t)-\Theta(t-T_d)$ usually takes very small values and change very slowly with time. It can be approximated by $\Phi(t)\approx\Phi(iT_{s}+t_m)~\textrm{for~}t\in[iT_{s}+T_{1},iT_{s}+T_{2}]$, where $t_m\triangleq(T_1+T_2)/2$, and then $\Phi(iT_{s}+t_m)$ can be considered as a zero-mean Gaussian r.v. with variance $4\pi\beta{T_d}$ (see Appendix~\ref{D_Phi_RV} for more detail). Therefore, the factor $\cos[\Phi(iT_{s}+t_m)]$ in (\ref{equi_DV_baseband}) can be moved outside the integral and it can be approximated by a truncated Taylor series expansion as follows
\begin{align}\label{Maclaurin_approx}
\cos\big[\Phi(iT_{s}+t_m)\big]&{\approx} 1-2\pi\beta{T_d}X^2,
\end{align}
where $X=\frac{\Phi(iT_{s}+t_m)}{2\sqrt{\pi\beta{T_d}}}$ is a zero-mean Gaussian r.v. with variance $1$. Substituting (\ref{Maclaurin_approx}) into (\ref{equi_DV_baseband}), we obtain
\begin{align}\label{equi_DV_final}
D&{\approx} {F_{a}}{F_{b}},
\end{align}
where $F_{a}\triangleq{\frac{1}{2}\int_{iT_{s}+T_{1}}^{iT_{s}+T_{2}}\breve{s}(t)\breve{s}(t-T_{d})dt}$, and $F_{b}\triangleq{1-2\pi\beta{T_d}Y}$, where $Y\triangleq{X^2}$ is a central chi-squared r.v. with degree of freedom $u=1$. It is noted that $F_a$ exactly corresponds to the DV of the baseband equivalent TRPC. According to the decision criterion (\ref{Decision}), the bit-error probability (BEP) for the passband TRPC conditioned on the channel realization $\mathbf{h}$ is then derived as
\begin{align}\label{Pe_passband_trpc}
P(e|\mathbf{h})&=\frac{1}{2}\Big\{(1-P_B^{+})-(1-2P_B^{+})P_{\phi}\Big\}\nonumber\\
&{+}\frac{1}{2}\Big\{(1-P_B^{-})-(1-2P_B^{-})P_{\phi}\Big\},
\end{align}
where the detailed derivation of (\ref{Pe_passband_trpc}) can be found in Appendix~\ref{D_Pe_passband}, and $\mathbf{h}=\{(\alpha_k,\tau_k)|k=0,\cdots,K-1\}$. $P_B^{+}$ and $P_B^{-}$ denotes the BEPs of the baseband TRPC conditioned on $b_i=+1$ and $b_i=-1$, respectively. $P_{\phi}=P(Y<\Omega)$ where $\Omega=1/(2\pi\beta{T_d})$. 

To calculate (\ref{Pe_passband_trpc}), the two conditional BEPs of the baseband TRPC can be obtained by using Gaussian approximation presented in \cite{Dong_trpc} as follows  
\begin{align}\label{Pe_baseband_pos}
P_B^{+}=Q\Big(\frac{\breve{m}_{D}^{+}}{\breve{\sigma}_{D}^{+}}\Big)
\end{align}
and 
\begin{align}\label{Pe_baseband_neg}
P_B^{-}=Q\Big(\frac{-\breve{m}_{D}^{-}}{\breve{\sigma}_{D}^{-}}\Big),
\end{align}
respectively, where $Q(x)\triangleq\frac{1}{\sqrt{2\pi}}\int_x^{\infty}e^{(-t^{2}/2)}dt$ denotes the Q-function\cite{Madhow_funmental_DCom}, $[\breve{m}_{D}^{+},(\breve{\sigma}_{D}^{+})^2]$ and $[\breve{m}_{D}^{-},(\breve{\sigma}_{D}^{-})^2]$ are the means and variances (see (9)--(11) in \cite{Dong_trpc} for more detail) of $\breve{D}$ conditioned on $b_i=+1$ and $b_i=-1$, respectively, where $\breve{D}=\int_{iT_{s}+T_{1}}^{iT_{s}+T_{2}}\breve{s}(t)\breve{s}(t-T_{d})dt$.
Moreover, the cumulative distribution function (CDF) of r.v. $Y$ can be calculated by
\begin{align}\label{P_phi}
P_{\phi}=P(Y<\Omega)=\frac{\gamma(\frac{u}{2},\frac{w}{2})}{\Gamma(\frac{u}{2})}\Big|_{(u=1,~w=\Omega)}=1-2Q\big(\sqrt{\Omega}\big),
\end{align}
where $\Gamma(x)\triangleq\int_{0}^{\infty}t^{x-1}e^{-t}dt$ and $\gamma(x,s)\triangleq\int_{0}^{s}t^{x-1}e^{-t}dt$ denote the Gamma function and the lower incomplete Gamma function respectively \cite{Gradshteyn_table_func}. Substituting (\ref{Pe_baseband_pos})--(\ref{P_phi}) into (\ref{Pe_passband_trpc}), the semi-analytical BEPs can be obtained for the passband TRPC in the presence of phase noise.

From (\ref{Pe_passband_trpc}), we see that $P_B^{+}$, $P_B^{-}$ and $P_{\phi}$ are the only contributors to the BEP of the passband TRPC. Moreover, according to (\ref{Pe_passband_trpc})--(\ref{P_phi}), for a given channel realization, two remarks representing the contributions of $\beta$ and $E_b/N_0$ on the BEP can be concluded as follows:

\emph{Remark I}: For a given $E_b/N_0$, both $P_B^{+}$ and $P_B^{-}$ remain unchanged. When $\beta$ increases, both $\Omega$ and $P_{\phi}$ decrease, and therefore this results in a larger BEP. Especially, we have $\lim_{\beta \to \infty}P_{\phi}=0$, $\lim_{\beta \to \infty}{P(e|\mathbf{h})}=1-0.5(P_B^{+}+P_B^{-})$; $\lim_{\beta \to 0}P_{\phi}=1$, and $\lim_{\beta \to 0}{P(e|\mathbf{h})}=0.5(P_B^{+}+P_B^{-})$. Therefore, the BEP of the passband TRPC will converge to that of the baseband TRPC if perfect oscillators are employed in the transceiver.

\emph{Remark II}: For a given $\beta$, both $\Omega$ and $P_{\phi}$ remain unchanged. When $E_b/N_0$ increases, both $P_B^{+}$ and $P_B^{-}$ become smaller. Especially, we have $\lim_{\rho \to \infty}P_B=0$ and $\lim_{\rho \to \infty}{P(e|\mathbf{h})}=1-P_{\phi}$, where $\rho\triangleq{E_b}/N_{0}$. That means in the presence of phase noise, there will be a BEP floor with the increase of ${E_b}/N_{0}$, and the larger $\beta$ is, the higher the BEP floor appears.

\section{Results and Discussions}\label{sec4}
In this section, some semi-analytical and simulation results are presented to evaluate the BER performance of the passband TRPC system in the presence of phase noise, constant carrier frequency offset and phase offset. The IEEE 802.15.3a CM1 and CM2 channels \cite{Foerster_Channel} are considered. All semi-analytical and simulation results are the average performance obtained over $100$ random channel realizations.

To obtain the semi-analytical BER curves, we modeled a baseband TRPC system, in which a training sequence with length $N_{t}$ symbols is used to estimate the channel-dependent parameters required in (\ref{Pe_baseband_pos}) and (\ref{Pe_baseband_neg}) for each channel realization. This training sequence consists of $N_{t}/2$ consecutive symbols ``$+1$'' and $N_{t}/2$ consecutive symbols ``$-1$''. Similar to the processing steps presented by (1)--(2), (5)--(7) in \cite{Dong_trpc}, the corresponding DV's can be obtained as $\breve{D}_{i}$ for $i=0,1,\cdots,N_{t}-1$. Then $[\breve{m}_{D}^{+},(\breve{\sigma}_{D}^{+})^2]$ and $[\breve{m}_{D}^{-},(\breve{\sigma}_{D}^{-})^2]$ can be estimated by calculating the means and the variances for the two sequences $\breve{D}^{+}$ and $\breve{D}^{-}$, where $\breve{D}^{+}\triangleq\{\breve{D}_{0},\breve{D}_{1},\cdots,\breve{D}_{N_{t}/2-1}\}$ and $\breve{D}_{i}^{-}\triangleq\{\breve{D}_{N_{t}/2},\breve{D}_{N_{t}/2+1},\cdots,\breve{D}_{N_{t}-1}\}$, respectively.
We found that satisfied results can be obtained when $N_{t}\geq{1024}$, and therefore we set $N_{t}={1024}$ in this paper.  

\Figure[t!](topskip=0pt, botskip=0pt, midskip=0pt)[scale=0.49]{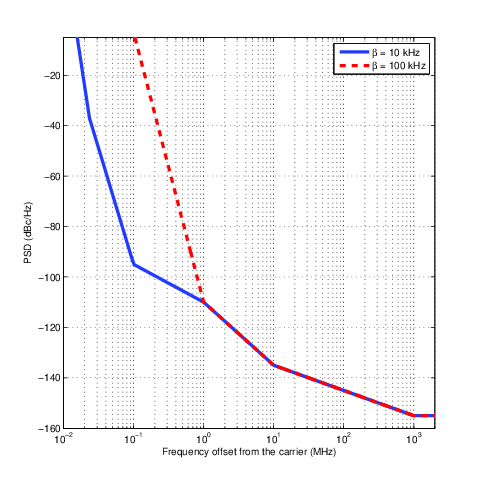}
{Simulated single side band (SSB) Lorentzian spectra of phase noise with $\beta=10$ kHz and $100$ kHz.\label{demo_phase_noise}} 

\subsection{Baseband Settings}
A root-raised-cosine (RRC) pulse with a roll-off factor of $0.25$ is used for the transmitter pulse shaping filter and the receiver LPFs. We assume the data rate is $R_b=1$ Mbps, $T_{s}=10^{3}$ ns, $N_{f}=4$, $T_{d}=T_{p}=2.02$ ns, and all other parameters are the same as those in \cite{Dong_trpc}.

\subsection{Simulated Model for Phase Noisy Oscillator}
Using the model given by (\ref{Carrier_phase_noisy})--(\ref{var_phase_noise}) and employing typical measured PSD profiles of phase noise presented in \cite{Razavi_phase_noise,Fard_PN_VCO,Huo_VCO,Huo_VCO_TRPC,Lu_ADPLL}, we can obtain a simulated phase noise via the following steps:

\emph{Step 1}: Define a frequency offset feature set and its corresponding PSD value set for the required phase noise model, where the frequency offset feature set includes the maximum frequency offset, $3$-dB bandwidth $\beta$, and several logarithmically spaced frequency offset points.

\emph{Step 2}: Based on the frequency offset feature set, generate an equally spaced frequency offset grid and perform interpolation of the PSD value set to form the corresponding PSD points.

\emph{Step 3}: Generate an AWGN vector with variance $1$ in frequency domain and weight it element by element with the interpolated PSD value set to obtain the simulated Lorentzian spectrum.

\emph{Step 4}: Perform Inverse Fast Fourier Transformation (IFFT) to the Lorentzian spectrum to obtain the required phase noise.

\Figure[t!](topskip=0pt, botskip=0pt, midskip=0pt)[scale=0.49]{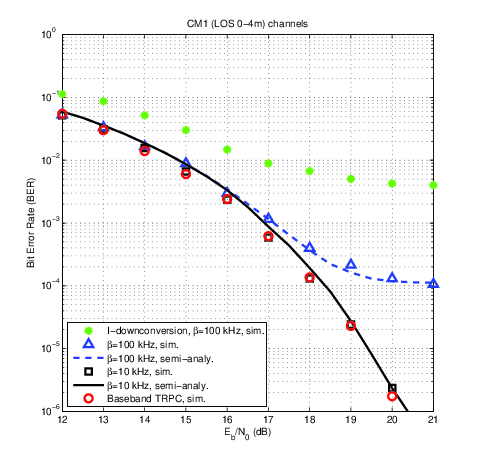}
{BER performance of the passband TRPC in the presence of phase noise with $\beta=10$ kHz, $100$ kHz and $200$ kHz in CM1 channels.\label{comp_semi_analytic_sim_cm1}} 

In the simulation, the nominal center frequency $f_{c}=3952$ MHz. Since the parameter $T_d$ is constant, we just need to use the $3$-dB bandwidth $\beta$ to characterize the level of phase noise. In low-power radio applications, the $\beta$-to-$f_c$ ratio can be as high as $1\times10^{-4}$ \cite{Wilcoxson_Master}, and therefore the maximum $\beta$ is set as $200$ kHz for the worst case of extremely noisy oscillators. Although $\beta$ usually can be as small as tens of Hertz for a high-quality oscillator in conventional radio systems, some of this stability is sacrificed to reduce the cost in low-power radio communications \cite{Wilcoxson_Master}. Therefore, we consider the minimum $\beta$ as $10$ kHz which represents a fairly conservative level for a practical oscillator operating at GHz bands for the passband TRPC. 
Moreover, if not specifically indicated, the constant carrier frequency offset and phase offset are modeled by the r.v.'s $\Delta{f}$ and $\phi$ uniformly distributed over $[-5,+5]$ MHz and $[0,2\pi)$, respectively. Fig.~\ref{demo_phase_noise} illustrates the simulated Lorentzian spectra of phase noise with $\beta=10$ kHz and $100$ kHz, respectively.

\Figure[t!](topskip=0pt, botskip=0pt, midskip=0pt)[scale=0.49]{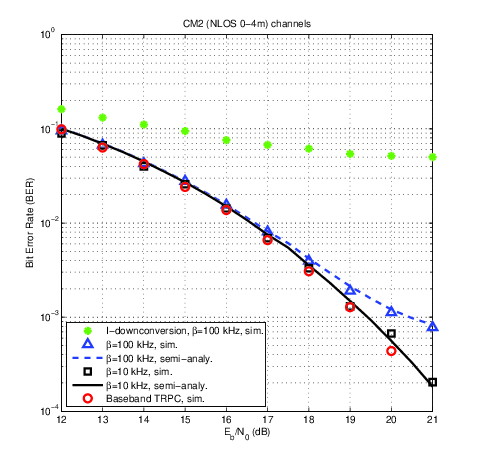}
{BER performance of the passband TRPC in the presence of phase noise with $\beta=10$ kHz and $100$ kHz in CM2 channels.\label{comp_semi_analytic_sim_cm2}} 

\subsection{Semi-analytical and Simulation Results}
Figs.~\ref{comp_semi_analytic_sim_cm1} and \ref{comp_semi_analytic_sim_cm2} present some simulation and semi-analytical results for the BER performance of the passband TRPC in the presence of phase noise with different $\beta$ values in CM1 and CM2 channels, respectively. The semi-analytical results were obtained by using (\ref{Pe_passband_trpc})--(\ref{P_phi}). We see that the simulation and semi-analytical results fit well for most $E_b/N_0$ values. However, compared to the simulation results, the semi-analytical results can be obtained much more quickly because the time-consuming statistical process of error bits can be avoided. 
\Figure[t!](topskip=0pt, botskip=0pt, midskip=0pt)[scale=0.54]{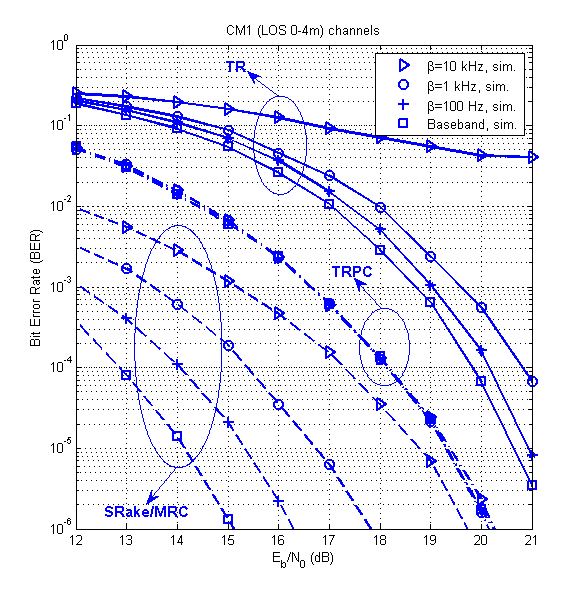}
{BER performance of the passband TRPC, the passband TR and the passband SRake/MRC-UWB with eight fingers, in the presence of phase noise with $\beta=100$ Hz, $1$ kHz and $10$ kHz, respectively, all in CM1 channels ($\beta=100$ Hz and $1$ kHz are only applicable to the passband TR and the passband SRake/MRC-UWB).\label{comp_sim_ctr_srake_cm1}}

From Figs.~\ref{comp_semi_analytic_sim_cm1} and \ref{comp_semi_analytic_sim_cm2}, we also see that when $\beta\leq{100}$ kHz in both CM1 and CM2 channels, the BER performance of the passband TRPC with noisy oscillators is very close to that of the passband TRPC with perfect oscillators (or that of the baseband TRPC presented in \cite{Liang_AD_TRPC}) for most $E_b/N_0$ values. However, there are power penalties of about $1.8$ dB with $\beta=100$ kHz at BER $=1\times{10^{-4}}$ for CM1 channels, and about $1$ dB with $\beta=100$ kHz at BER $=1\times{10^{-3}}$ for CM2 channels, respectively. Therefore, in the implementation of passband TRPC systems, suitable oscillators need to be employed according to the requirements of different applications.
 
At the same time, to show the effect of I-Q downconversion in terms of constant carrier frequency offset and phase offset cancellations, Figs.~\ref{comp_semi_analytic_sim_cm1} and \ref{comp_semi_analytic_sim_cm2} also present the BER performance of the passband TRPC using single-branch downconversion (I- or Q-downconversion) in CM1 and CM2 channels, respectively. As expected, we see that the I-Q downconversion outperforms the I-downconversion significantly and single-branch downconversion should not be used without recovery of carrier phase. For example, for a given $\beta=100$ kHz, the BER floors of the I-Q downconversion are lower than those of the I-downconversion by one or two orders of magnitude in both CM1 and CM2 channels. According to the discussions below (\ref{combined_equi_baseband}), we see that these performance gaps result from the constant carrier frequency offset $\Delta{f}$ and the phase offset $\phi$. 

In order to obtain a comprehensive comparison with other impulse radio technologies, the passband TR system, and the passband coherent UWB system using selective Rake (SRake) receiver with maximum ratio combing (MRC), are also considered to evaluate the BER performance in the presence of phase noise (see Appendix~\ref{Model_CTR_SRAKE} for more detail).

Figs.~\ref{comp_sim_ctr_srake_cm1} and \ref{comp_sim_ctr_srake_cm2} present the simulation results for the BER performance of the passband TRPC, the passband TR and the passband SRake/MRC-UWB systems, in CM1 and CM2 channels, respectively. Due to the limited space, we only consider the phase noise and the phase offset and assume that there is no constant carrier frequency offset in the passband TR and the passband SRake/MRC-UWB. From Figs. \ref{comp_sim_ctr_srake_cm1} and \ref{comp_sim_ctr_srake_cm2} we see that when $\beta$ increases from $100$ Hz to $10$ kHz, which is a much more relaxed case compared with that of the passband TRPC, there are still power penalties of about $4.5$ dB for the SRake/MRC-UWB at BER $=1\times{10^{-6}}$ for CM1 channels, and about $6$ dB at BER $=2\times{10^{-4}}$ for CM2 channels, respectively. For the passband TR, although the power penalties are less than $3$ dB when $\beta\leq{1}$ kHz, high BER floors appear when $\beta$ reaches $10$ kHz. In contrast, there is no appreciable performance loss for the passband TRPC in the presence of phase noise with $\beta=10$ kHz, indicating TRPC is much more immune to phase noise than conventional TR and SRake/MRC-UWB systems.

\Figure[t!](topskip=0pt, botskip=0pt, midskip=0pt)[scale=0.54]{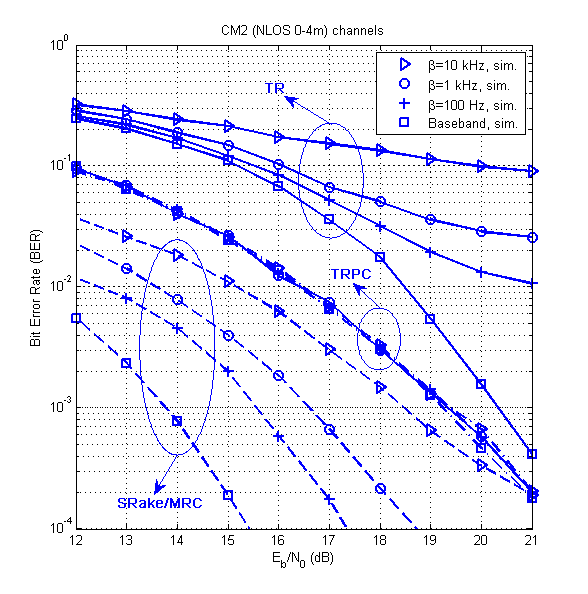}
{BER performance of the passband TRPC, the passband TR and the passband SRake/MRC-UWB with sixteen fingers, in the presence of phase noise with $\beta=100$ Hz, $1$ kHz and $10$ kHz, respectively, all in CM2 channels ($\beta=100$ Hz and $1$ kHz are only applicable to the passband TR and the passband SRake/MRC-UWB).\label{comp_sim_ctr_srake_cm2}}

\subsection{Discussions on Spectral Efficiency, Hardware Complexity and Cost}
Based on the discussions below (\ref{passband_trpc}) and the system models presented in Appendix~\ref{Model_CTR_SRAKE}, we see that the I-Q upconversion with different I-Q input can be employed in the passband SRake/MRC-UWB. For the passband TRPC and the passband TR, the use of the I-Q upconversion, or equivalently, the I- (or Q-) upconversion results in a fifty-percent penalty in terms of spectral efficiency. However, this cost is acceptable for the passband TRPC if advanced spectrum-sharing technologies are employed. For the passband SRake/MRC-UWB, apart from channel estimation algorithms for coherent detection, efficient methods are also needed to combat frequency or timing jitter caused by phase noise, and accordingly, the dramatically increased complexity will inevitably become a major challenge to hardware implementation. Moreover, according to the results presented in Figs.~\ref{comp_sim_ctr_srake_cm1} and \ref{comp_sim_ctr_srake_cm2}, we can conclude that in order to avoid significant BER performance penalties, costly ultralow phase noise oscillators are required in the passband TR and the passband SRake/MRC-UWB. In comparison, low cost oscillators even without phase-locked loop are sufficient for the passband TRPC due to its robust BER performance against phase noise.  

\section{Conclusion}\label{sec5}
In this paper, we have presented the design and performance analysis of practical passband TRPC-UWB systems in the presence of phase noise. We have shown that the constant carrier frequency offset and the phase offset can be successfully cancelled with the passband transmitter and the noncoherent receiver employed in this paper. Semi-analytical and simulation results have verified the accuracy of our analyses and therefore they can be employed in the implementation of passband TRPC-UWB systems. Our results have also demonstrated that compared to the passband TR and the passband SRake/MRC-UWB systems, the passband TRPC has robust system performance against phase noise, constant carrier frequency offset, and phase offset, and therefore it is a promising candidate for low cost, low power consumption, and low complexity applications. 

\appendices

\appendices
\section{Discussions of the approximation}\label{LTI_model}
According to (\ref{phase_noise}), we see that $\theta(t)$ tends to be a slowly time-varying Gaussian random process due to the integration operation. Measured results show that for most feasible oscillators operating at the UWB frequency bands, the PSD at $1$ MHz offset is around $-120$ dBc/Hz, and especially, the quantity at $10$ MHz offset is less than $-130$ dBc/Hz\cite{Fard_PN_VCO,Huo_VCO,Huo_VCO_TRPC,Lu_ADPLL}. Therefore, for most practical oscillators used in low cost, low power consumption applications, we consider that the maximum frequency offset caused by phase noise ranges from $1$ MHz to $10$ MHz. That means the output phase noise $\theta(t)$ has a coherent time which ranges from $1\times10^2$ to $1\times10^3$ ns. For most IEEE 802.15.3a indoor channels, the mean excess delay ranges several nanoseconds from tens of nanoseconds \cite{Foerster_Channel}. Therefore, we consider $\theta(t-\tau_{k})\approx\theta(t)~\textrm{for}~k=0,\cdots,K-1$.

\section{Derivation of the variance}\label{D_Phi_RV}
Without loss of generality, let $i=0$ and we have $\Phi(t_m)=\Theta(t_m)-\Theta(t_m-T_d)$, where $\Theta(t)=\theta_{tx}(t)-\theta_{rx}(t)$. According to (\ref{phase_noise}) and (\ref{var_phase_noise}), $\theta_{tx}(t)$ and $\theta_{rx}(t)$ can be considered as identically distributed (i.d.) zero-mean Gaussian r.v.'s with variance $2\pi\beta{t}$. Similarly, $\Theta(t_m)$ can be approximated as an Gaussian r.v. with variance $4\pi\beta{t}$. Therefore, $\Phi(t_m)$ can be modeled as a zero-mean Gaussian r.v. with variance $\sigma_{\Phi}^2$, where $\sigma_{\Phi}^2$ can be calculated as follows:
\begin{align}\label{sigma_phi}
\sigma_{\Phi}^2&{=} Var\big\{\Theta(t_m)-\Theta(t_m-T_d)\big\}\nonumber\\
&{=} Var\big\{\Theta(t_m)\big\}+Var\big\{\Theta(t_m-T_d)\big\}\nonumber\\
&{-} 2E\big\{\Theta(t_m)\Theta(t_m-T_d)\big\}\nonumber\\
&{=} 8\pi{\beta}{t_m}-4\pi{\beta}{T_d}-4R_{\theta}(t_m,T_d),
\end{align}
where $R_{\theta}(t,\tau)\triangleq{E\big[\theta(t)\theta(t-\tau)\big]}$ denotes the autocorrelation function of $\theta(t)$. Using (\ref{phase_noise}) and (\ref{var_phase_noise}), we obtain $R_{\theta}(t_m,T_{d})$ as $2\pi\beta{(t_m-T_d)}$, and substituting it into (\ref{sigma_phi}), we have $\sigma_{\Phi}^2\approx{4\pi{\beta}{T_d}}$.   

\section{Derivation of the BEP for Passband TRPC}\label{D_Pe_passband}
Using (\ref{Decision}) and (\ref{equi_DV_final}), we have
\begin{align}\label{Deriv_Pe_passband_trpc}
P(e|\mathbf{h})&{=} \frac{1}{2}P(D<0|b_{i}=+1)+\frac{1}{2}P(D>0|b_{i}=-1)\nonumber\\
&{=} \frac{1}{2}P_{1}+\frac{1}{2}P_{2},
\end{align}
where $P_{1}=P(F_a\cdot{F_b}<0|b_{i}=+1)$ and $P_{2}=P(F_a\cdot{F_b}>0|b_{i}=-1)$ can be derived as
\begin{align}\label{cond_pe_pos}
P_1&{=} (1-P_B^{+}){P(Y>\Omega)}+{P_B^{+}}{P(Y<\Omega)}\nonumber\\
&{=} (1-P_B^{+})-(1-2{P_B^{+}})P_{\phi}
\end{align}
and
\begin{align}\label{cond_pe_neg}
P_2&{=} (1-P_B^{-}){P(Y>\Omega)}+{P_B^{-}}{P(Y<\Omega)}\nonumber\\
&{=} (1-P_B^{-})-(1-2{P_B^{-}})P_{\phi},
\end{align}
respectively, where $F_{a}\triangleq{\frac{1}{2}\int_{iT_{s}+T_{1}}^{iT_{s}+T_{2}}\breve{s}(t)\breve{s}(t-T_{d})dt}$, $F_{b}\triangleq{1-2\pi\beta{T_d}Y}$, $\Omega=1/(2\pi\beta{T_d})$. Substituting (\ref{cond_pe_pos}) and (\ref{cond_pe_neg}) into (\ref{Deriv_Pe_passband_trpc}), we obtain (\ref{Pe_passband_trpc}).

\section{System models for passband TR and passband SRake/MRC-UWB}\label{Model_CTR_SRAKE}
For the passband TR system, the upconversion operation can be performed in the I-Q upconversion, or equivalently, in the form of I-upconversion (or Q-upconversion). Without loss of generality, we consider the I-Q upconversion for the passband TR system, where the system models are the same as those of the passband TRPC presented in Fig. \ref{block_diagram}, except that the baseband signal is replaced by the TR signaling as below:
\begin{align}\label{baseband_ctr}
s_{tr}(t)&= \sqrt{\frac{E_b}{2N_f}}
\sum_{m=0}^{N_f-1}[g(t-iT_s-mT_f)\nonumber\\
&{+} b_i\cdot{g(t-iT_s-mT_f-T_d^{'})}],
\end{align}
where $T_d^{'}\geq\tau_{max}+T_p$, $T_{f}\geq2T_d^{'}$ and $T_{s}\geq{N_{f}}T_{f}$, and other notations are the same as those in (\ref{baseband_trpc}). 

For the passband SRake/MRC-UWB system, the bipolar baseband signal can be modeled as
\begin{align}\label{baseband_bpsk}
s{'}(t)=\sqrt{\frac{E_b}{2N_f}}
\sum_{m=0}^{2N_f-1}b_i\cdot{g(t-iT_s-mT^{'}_f)},
\end{align}
where $T^{'}_{f}=T_{f}/2$, and other notations are the same as those in (\ref{baseband_trpc}) and (\ref{baseband_ctr}). For the passband SRake/MRC-UWB system, the I-Q upconversion with different I-Q input can be employed due to the use of coherent detection techniques. For simplicity, only the I-upconversion is considered in this paper and therefore the transmitted passband signal is
\begin{equation}\label{passband_bpsk}
s{'}_T(t)=\sqrt{2}s{'}(t)\cos\left[2\pi{f_c}t+\theta_{tx}(t)\right].
\end{equation}
After $s{'}_T(t)$ passes through the multipath channel and is corrupted by AWGN, the received passband signal is
\begin{align}\label{received_bpsk}
r{'}(t)=\sum_{k=0}^{K-1} \alpha_k s{'}_T(t-\tau_k)+n(t).
\end{align}
Via downconversion, the received signal becomes     
\begin{align}\label{downconvert_bpsk}
\tilde{r}{'}(t)=r{'}(t)\cos\left[2\pi({f_c}+\Delta{f})t+\theta_{rx}(t)+\phi\right].
\end{align}
By LPF-filtering, the baseband received signal $r_{B}(t)$ can also be obtained. Assuming that the receiver can perfectly estimate the channel state information and following the processing steps given by (5)--(7) and (9) in \cite{Choi_Performance}, the MRC combiner's output $Z_{i}$ can be obtained as the decision variable $D{'}$ for the $i$-th symbol. Finally, a symbol decision can be made by using (\ref{Decision}).

\bibliographystyle{IEEEtran}
\bibliography{IEEEabrv,reference}

\begin{IEEEbiography}[{\includegraphics[width=1in,height=1.25in,keepaspectratio]{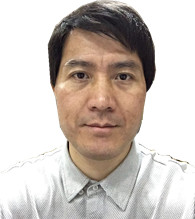}}]{Zhonghua Liang}(S'07--M'08) received the B.Sc. degree in radio engineering, the M.Sc. and Ph.D. degrees in information and communication engineering from Xi'an Jiaotong University, Xi'an, China, in 1996, 2002, and 2007, respectively.

From July 1996 to August 1999, he was with the Guilin Institute of Optical Communications (GIOC), Guilin, China, where he was a system engineer of optical transmission systems. Between 2008 and 2009, he worked as a postdoctoral fellow in the Department of Electrical and Computer Engineering, University of Victoria, Victoria, BC, Canada. Since 2010, he has been with the School of Information Engineering, Chang'an University, Xi'an, China, where he is currently an associate professor. His research interests include mobile communication systems, ultrawideband radio, and signal processing for wireless communication systems.
\end{IEEEbiography}

\begin{IEEEbiography}[{\includegraphics[width=1in,height=1.25in,clip,keepaspectratio]{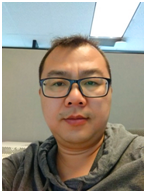}}]{Guowei Zhang} was born in Anyang, China. He received his bachelor degrees in English and Computer science from the University of science and Technology of China, Hefei, China in 2004 and M.S degree in Optics from Chinese Academy of Sciences in 2008. He is currently a hardware engineer working in Vancouver, Canada.

Mr. Zhang's research interest covers design and analysis of optimized UWB receiving mechanism and implementation of UWB wireless communication system, using RF component and FPGA, INTEL CPU platform bring up and power solutions. Mr. Zhang received graduate fellowship throughout years at University of Victoria. He also had three patents related to cell phone devices.
\end{IEEEbiography}

\begin{IEEEbiography}[{\includegraphics[width=1in,height=1.25in,keepaspectratio]{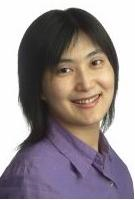}}]{Xiaodai Dong}(S'97--M'00--SM'09) received the B.Sc. degree in information and control engineering from Xi'an Jiaotong University, China, in 1992, the M.Sc. degree in electrical engineering from the National University of Singapore in 1995, and the Ph.D. degree in electrical and computer engineering from Queen's University, Kingston, ON, Canada, in 2000.

From 1999 to 2002, she was with Nortel Networks, Ottawa, ON, Canada, where she was involved in the base transceiver design of the third-generation (3G) mobile communication systems. From 2002 to 2004, she was an Assistant Professor with the Department of Electrical and Computer Engineering, University of Alberta, Edmonton, AB, Canada. She was a Canada Research Chair (Tier II) from 2005 to 2015. Since 2005, she has been with the University of Victoria, Victoria, Canada, where she is currently a Professor with the Department of Electrical and Computer Engineering. Her research interests include wireless communications, radio propagation, ultra-wideband radio, machine-to-machine communications, wireless security, eHealth, smart grid, nanocommunications, and signal processing for communication applications.

Dr. Dong served as an Editor of \textsc{the IEEE Transactions on Wireless Communications} from 2009 to 2014, \textsc{the IEEE Transactions on Communications} from 2001 to 2007, and \textsc{the Journal of Communications and Networks} from 2006 to 2015. She is currently an Editor of \textsc{the IEEE Transactions on Vehicular Technology}.
\end{IEEEbiography}

\begin{IEEEbiography}[{\includegraphics[width=1in,height=1.25in,keepaspectratio]{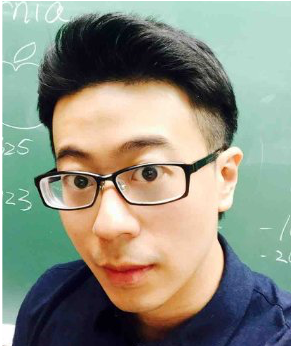}}]{Yiming Huo}(S'08--M'18) received B.Eng degree in information engineering from Southeast University, China, in 2006, MSc. degree in system-on-chip (SOC) from Lund University, Sweden, in 2010, and the Ph.D. in electrical engineering at University of Victoria, Canada, in 2017. He is currently a Research Associate with the Electrical and Computer Engineering Department, University of Victoria, BC, Canada. His current research interests are in the physical layer design of 5G and IoT wireless communications systems.

From 2009 to 2010, he stayed in Ericsson and ST-Ericsson where he accomplished his master thesis on multi-mode, wideband CMOS VCOs for cellular systems. Between 2010 and 2011, Mr. Huo worked in Chinese Academy of Sciences (CAS) as research associate. From 2011 to 2012, he worked in STMicroelectronics as RF engineer for developing digital video broadcasting (DVB) systems. From 2015 to 2016, he spent a year conducting RF hardware and cellular design for various projects in Apple Inc., Cupertino, California, as a Ph.D intern.  

Dr. Huo was a recipient of the Best Student Paper Award of the 2016 IEEE ICUWB, the Excellent Student Paper Award of the 2014 IEEE ICSICT, and the Bronze Leaf certificate of the 2010 IEEE PrimeAsia. He also received the University of Victoria Fellowship from 2012 to 2013, and ISSCC-STGA from the IEEE Solid-State Circuits Society (SSCS) in 2017. He is a member of the SSCS, the CASS, the MTT-S, and the ComSoc. He served as the Program Committee member of the IEEE ICUWB 2017, TPC member of IEEE VTC-Spring 2018, and a technical reviewer for multiple IEEE conferences and journals.
\end{IEEEbiography}
\EOD

\end{document}